\documentclass{aa}  
\usepackage{graphicx}
\usepackage{txfonts}
\begin{document}

\title{Adaptive smoothing lengths in SPH}

\author{R. E. Attwood\inst{1}
\and S. P. Goodwin\inst{2}
\and A. P. Whitworth\inst{1}}

\offprints{R. E. Attwood}

\institute{School of Physics \& Astronomy, Cardiff University, Queens Buildings, The Parade, Cardiff CF24 3AA, Wales, UK\\
\email{Rhianne.Attwood@astro.cf.ac.uk}\\
\email{Anthony.Whitworth@astro.cf.ac.uk}
\and Department of Physics \& Astronomy, University of Sheffield, Hicks Building, Housfield Road, Sheffield S3 7RH, UK\\
\email{S.Goodwin@Sheffield.ac.uk}}

\date{Received 20 October 2006 ; accepted 18 December 2006}

\abstract
{There is a need to improve the fidelity of SPH simulations of self-gravitating gas dynamics.}
{We remind users of SPH that, if smoothing lengths are adjusted so as to keep the number of neighbours, ${\cal N}$, in the range ${\cal N}_{_{\rm NEIB}}\pm\Delta{\cal N}_{_{\rm NEIB}}$, the tolerance, $\Delta{\cal N}_{_{\rm NEIB}}$, should be set to zero, as first noted by Nelson \& Papaloizou. We point out that this is a very straightforward and computationally inexpensive constraint to implement.}
{We demonstrate this by simulating acoustic oscillations of a self-gravitating isentropic monatomic gas-sphere (cf. Lucy), using ${\cal N}_{_{\rm TOT}}\sim6,000$ particles and ${\cal N}_{_{\rm NEIB}}=50$.}
{We show that there is a marked reduction in the rates of numerical dissipation and diffusion as $\Delta{\cal N}_{_{\rm NEIB}}$ is reduced from 10 to zero. Moreover this reduction incurs a very small computational overhead.}
{We propose that this should become a standard test for codes used in simulating star formation. It is a highly relevant test, because pressure waves generated by the switch from approximate isothermality to approximate adiabaticity play a critical role in the fragmentation of collapsing prestellar cores. Since many SPH simulations in the literature use ${\cal N}_{_{\rm NEIB}}=50$ and $\Delta{\cal N}_{_{\rm NEIB}}\geq10$, their results must be viewed with caution.}

\keywords{Hydrodynamics - Methods: numerical - Stars: oscillations}

\maketitle

\section{Introduction}

Numerical simulations play an increasingly important role in the study of star formation, and of other non-linear phenomena involving self-gravitating gas dynamics, for example the development of cosmological structure, galaxy formation and stellar collisions. Since these problems involve complex three-dimensional configurations and/or large ranges of density, Smoothed Particle Hydrodynamics (SPH; e.g. Monaghan 1992) is well suited to such simulations. The principal alternative is to use an Adaptive Mesh Refinement (AMR) Finite Difference code (e.g. Truelove et al. 1998), although this is in general more expensive computationally. However, ultimately neither method is useful, unless it can be shown that the results are converged, and are not compromised by numerical artefact. This requires test problems with known analytic, or semi-analytic, solutions, which involve the same basic physical phenomena as occur during star formation.

We propose such a test here, namely acoustic oscillations of a self-gravitating, isentropic, monatomic gas-sphere in the fundamental radial mode. This test was originally performed by Lucy (1977) in his seminal paper introducing SPH, and therefore we shall refer to it as the Lucy test. It has been performed subsequently (e.g. Steinmetz \& M\"uller 1993; Nelson \& Papaloizou 1994), but infrequently. It is an appropriate test because it measures (i) the level of dissipation associated with artificial viscosity, in the absence of shocks; (ii) the extent to which transients, due to the discrete nature of particles (or cells), remove energy from genuine modes and transfer it to other spurious modes (i.e. numerical diffusion); (iii) the ability of the code to model acoustic oscillations, and in particular adiabatic bounces; and (iv) the ability of the code to deal with free (or nearly free) boundaries.

Point (iii) is particularly important because it seems that collapsing prestellar cores are most prone to fragment at the stage when the gas switches from being approximately isothermal to being approximately adiabatic (e.g. Boss et al. 2000). Fragmentation at this juncture is presumably due to interactions between the complex system of pressure waves which is generated by adiabatic bounces in a converging but disordered inflow, and it is therefore essential that spurious waves are not being generated. In this context it is worth noting that the isentropic assumption is not strictly the same as adiabaticity, and is made here for analytic convenience rather than realism. In simulations where shocks develop, artificial viscosity must be incorporated, in which case, {\it either} the resulting energy dissipation must be included in the energy equation, {\it or} -- if the thermal timescale is sufficiently short, and the main concern is not with the detailed structure of the shock front -- a barotropic equation of state may be invoked. There are no shocks in the present simulation, and therefore the rate of energy dissipation due to artificial viscosity is low, but not negligible. The isentropic assumption then simply implies that the small amount of energy dissipated by artificial viscosity is spirited away by some unspecificed process.

It is also appropriate to point out that, although the gas in prestellar cores is largely molecular, it behaves as a monatomic gas (i.e. effective adiabatic exponent $\gamma=5/3$) until the temperature rises above $\sim 100\,{\rm K}$. At lower temperatures the rotational degrees of freedom are not significantly excited, since para-H$_{_2}$ has its first excited level ($J=2$) at $k_{_{\rm B}}(512\,{\rm K})$, and ortho-H$_{_2}$ has its first excited level ($J=3$) at $k_{_{\rm B}}(854\,{\rm K})$ (e.g. Black \& Bodenheimer 1975). Neglect of this fact can lead to artificially enhanced fragmentation of collapsing prestellar cores, since with $\gamma=5/3$ the Jeans mass increases quite rapidly with increasing density in the adiabatic regime ($M_{_{\rm JEANS}}\propto\rho^{1/2}$), whereas with $\gamma=7/5$ it increases much more slowly ($M_{_{\rm JEANS}}\propto\rho^{1/10}$).

In Section \ref{SEC:NUM}, we describe the numerical code used and how the test is initiated. In Section \ref{SEC:RES} we present and discuss the results. In Section \ref{SEC:CON} we summarise our conclusions.

\section{Numerical details}\label{SEC:NUM}

The code we have used -- {\sc dragon} (Goodwin et al. 2004) -- is a standard SPH code, using a second-order Runge-Kutta integration scheme, multiple particle timesteps, adaptive smoothing lengths, time-dependent artificial viscosity (Morris \& Monaghan 1997), and an octal spatial-decomposition tree (Barnes \& Hut 1986) for calculating gravitational accelerations and constructing neighbour lists. In calculating gravitational accelerations, tree-cells are opened if they present an angle greater than $\theta_{_{\rm CRIT}}=0.5$, and quadrupole moments are included. The smoothing length of a particle is adjusted so that its neighbour list contains ${\cal N} = {\cal N}_{_{\rm NEIB}}\pm\Delta{\cal N}_{_{\rm NEIB}}$ other particles\footnote{Some SPH codes (e.g. Price \& Monaghan 2004) do not specify the number of neighbours, but instead, for each particle, iterate around a loop,
\begin{eqnarray} \nonumber
h_{_i}&=&h_{_0}\,\left[\frac{m_{_i}}{\rho_{_i}}\right]^{1/3}\,,\\\nonumber
\rho_{_i}&=&\sum_{_j}\left\{\frac{m_{_j}}{h_{_i}^3}\,W\left(\frac{|{\bf r}_{_j}-{\bf r}_{_i}|}{h_{_i}}\right)\right\}\,,
\end{eqnarray}
until fractional changes in $\rho_{_i}$ drop below a user-defined tolerance, $\epsilon$. Here $h_{_0}$ is a constant of order unity, and the summation is over all neighbours $j$ for which $|{\bf r}_{_j}-{\bf r}_{_i}|<2\,h_{_i}$ (i.e. all particles within the smoothing kernel of particle $i$). Provided $\epsilon$ is sufficiently small, this is statistically equivalent to setting ${\cal N}_{_{\rm NEIB}}=32\pi h_{_0}^3/3$ and $\Delta{\cal N}_{_{\rm NEIB}}=0$.}.

\begin{figure*}\label{FIG:DISPLACEMENT}\centering
\includegraphics[angle=-90,width=\textwidth]{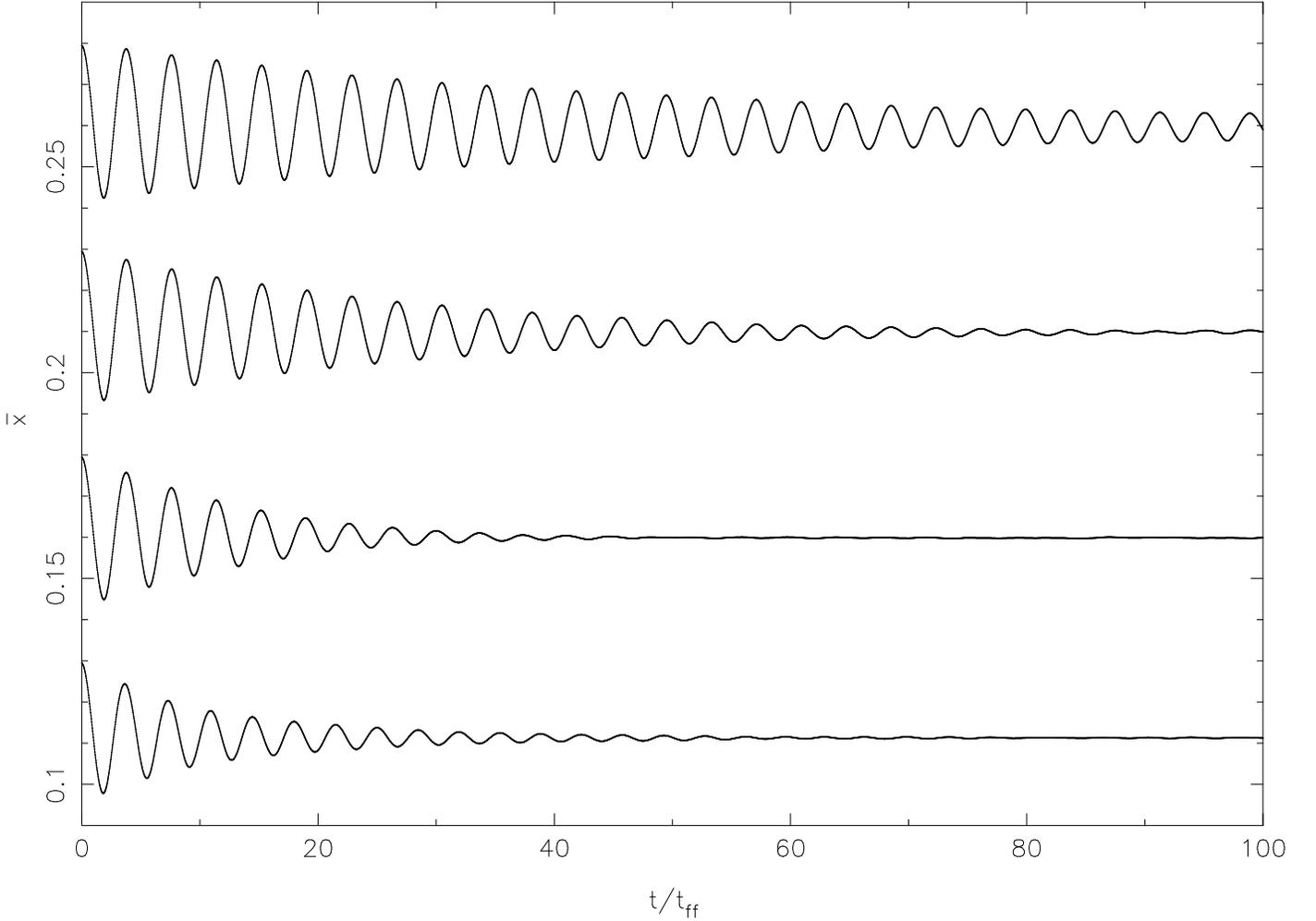}
\caption{The mean $x$-displacement, $\bar{x}$ (where the mean is over
  all particles, with equal weighting), against time, $t$ (in units of
  the freefall time in the equilibrium state, $t_{_{\rm FF}}$). The
  results are displaced vertically by $\Delta x$ to fit them all on
  one plot. Reading from the top, (a) $\Delta{\cal N}_{_{\rm NEIB}}\;=0,\,\Delta x = 0;\;$ (b) $\Delta{\cal N}_{_{\rm NEIB}}\;=2,\,\Delta x = -0.05;\;$ (c) $\Delta{\cal N}_{_{\rm NEIB}}\;=5,\,\Delta x = -0.10;\;$ (d) $\Delta{\cal N}_{_{\rm NEIB}}\;=10,\,\Delta x = -0.15\;.$}
\end{figure*}

As an illustration of the Lucy test, we demonstrate how much the
performance of this code improves as $\Delta {\cal N}_{_{\rm NEIB}}$
is reduced, and how marginal the extra computational cost is. The
equilibrium isentropic sphere is set up by placing equal-mass
particles on an hexagonal close-packed array within a sphere of radius
unity; then stretching this uniform-density sphere radially to
reproduce the density profile of a polytrope with exponent $5/3$ (or
equivalently index $3/2$); and finally relaxing the system by evolving
the particle positions using the SPH code, until the net kinetic
energy falls to a very small value (relative to the magnitude of the
gravitational potential energy). In the results presented here the
sphere is represented by ${\cal N}_{_{\rm TOT}}=5,895$ particles, and
we set ${\cal N}_{_{\rm NEIB}}=50$. The parameters in the
time-dependent artificial viscosity (Morris \& Monaghan 1997) are
$\alpha_{_\star}=0.1$, ${\cal C}_{_1}=0.2$, $\beta=2\alpha$. To excite
the fundamental mode, each particle is displaced radially from its
equilibrium radius $r$ to a new radius $r'=r\left[1+A\xi(r/R)\right]$,
and then released from rest. Here $\xi(s)$ ($0\leq s \leq
1$) is the eigenfunction of the fundamental radial mode\footnote{The
  eigenfunction, $\xi(s)$, for the fundamental radial mode of a self-gravitating isentropic monatomic gas-sphere was very kindly supplied, in the form of a dense look-up table, by Alfred Gautschy.}. $A$ is the initial amplitude of the oscillation, which we set to $A=0.1$.

\begin{figure*}\label{FIG:ENERGY}\centering
\includegraphics[angle=-90,width=\textwidth]{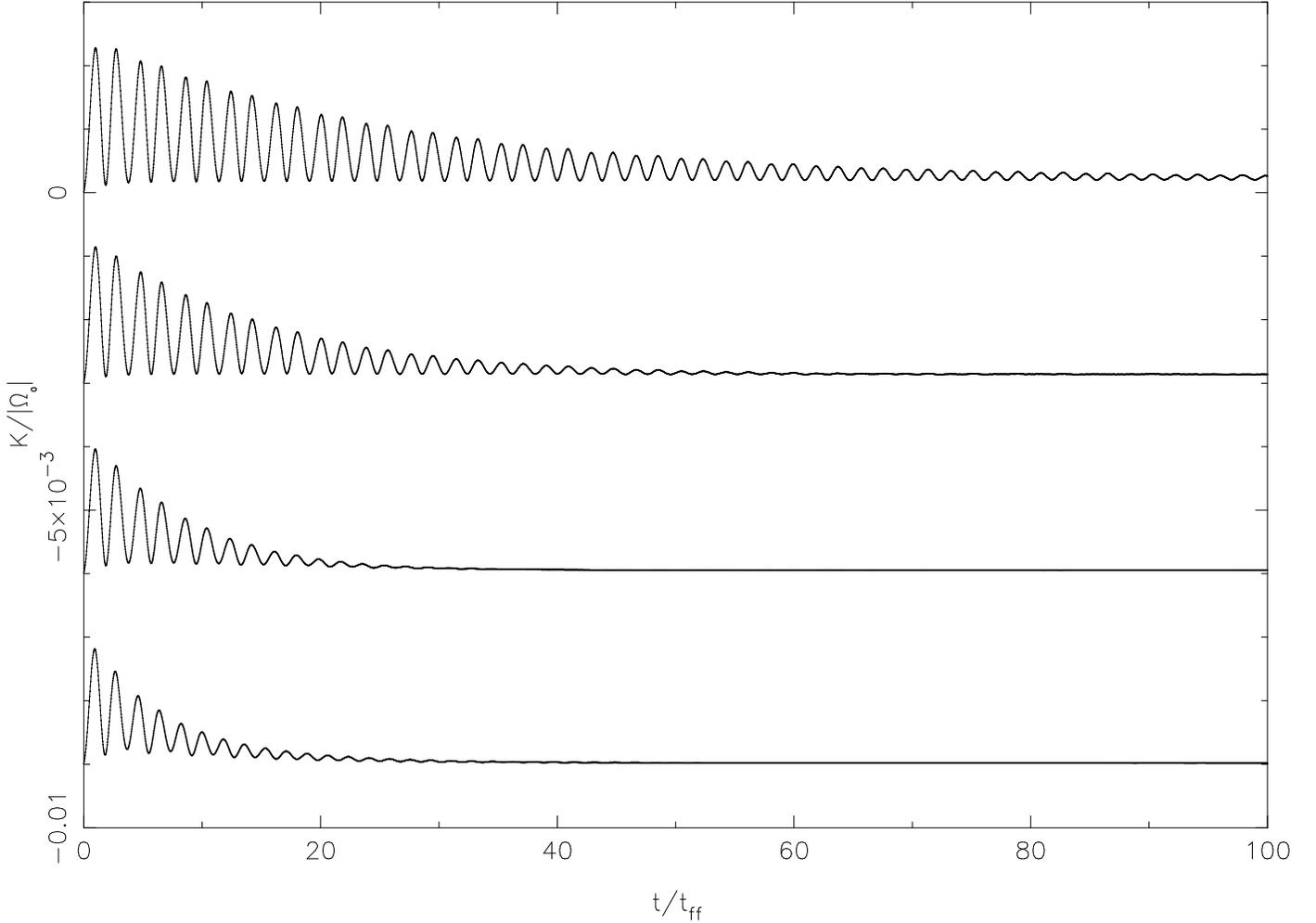}
\caption{The total kinetic energy, ${\cal K}$ (normalised to the
  magitude of the self-gravitational potential energy in the
  equilibrium state, $|\Omega_{_0}|$), against time, $t$ (in units of
  the freefall time in the equilibrium state, $t_{_{\rm FF}}$). The
  results are displaced vertically by $\Delta {\cal K}$ to fit them
  all on one plot. Reading from the top, (a) $\Delta{\cal N}_{_{\rm NEIB}}\;=0,\,\Delta {\cal K} = 0;\;$ (b) $\Delta{\cal N}_{_{\rm NEIB}}\;=2,\,\Delta {\cal K} = -0.003;\;$ (c) $\Delta{\cal N}_{_{\rm NEIB}}\;=5,\,\Delta {\cal K} = -0.006;\;$ (d) $\Delta{\cal N}_{_{\rm NEIB}}\;=10,\,\Delta {\cal K} = -0.009\;.$}
\end{figure*}

\section{Results and Discussion}\label{SEC:RES}

Figures 1 and 2 show oscillations simulated with $\Delta {\cal N}_{_{\rm NEIB}}=0,\,2,\,5,\;{\rm and}\;10$. In Fig. 1, the quantity plotted is the mean $x$-displacement
\begin{eqnarray}
\bar{x}&=&\frac{1}{{\cal N}_{_{\rm {\bf TOT}}}}\,\sum_{_i}\left\{|x_{_i}|\right\}
\end{eqnarray}
(the mean being taken over all the particles, with equal weighting),
as a function of time (normalised to the central freefall time, $t_{_{\rm FF}}$, in the unperturbed equilbrium state). In Fig. 2, the quantity plotted is the total kinetic energy, normalised to the magnitude of the gravitational potential energy in the equilibrium state (i.e. ${\cal K}/|\Omega_{_0}|$, where ${\cal K}$ is the total kinetic energy, $\Omega$ is the total gravitational potential energy, and subscript 0 refers to values in the unperturbed equilibrium state), again as a function of time (normalised to $t_{_{\rm FF}}$). Some decay parameters are recorded in Table 1.

\begin{table}
\caption{Decay of the fundamental mode. The first column gives
  $\Delta{\cal N}_{_{\rm NEIB}}$. The second column gives the
  e-folding time for the amplitude of the fundamental mode, $T_{_A}$,
  as a multiple of its period, $P_{_0}$. The third column gives the
  net oscillation energy left after 10 periods, ${\cal E}_{_{10}}$, as
  a fraction of the initial oscillation energy, ${\cal E}_{_0}$. The
  fourth column gives the simulation time to evolve the oscillating
  gas-sphere for 10 periods on eight 2.2Ghz Opteron CPUs each with 8GB memory, $t_{_{10}}$.}
\label{TAB:RESULTS}
\centering
\begin{tabular}{rrrr}\hline\hline
&&&\\
$\Delta{\cal N}_{_{\rm NEIB}}$&\hspace{1.0cm}$T_{_A}/P_{_0}$&\hspace{1.0cm}${\cal E}_{_{10}}/{\cal E}_{_0}$&\hspace{1.0cm}$t_{_{10}}/{\rm s}\,$\\
&&&\\\hline
&&&\\
0$\;\;\;\;$&13.6$\;\,$&0.955$\;$&9545\\
2$\;\;\;\;$&7.6$\;\,$&0.939$\;$&9403\\
5$\;\;\;\;$&3.8$\;\,$&0.927$\;$&8996\\
10$\;\;\;\;$&3.1$\;\,$&0.925$\;$&8164\\
&&&\\\hline
\end{tabular}
\end{table}

In addition to the fundamental mode, some additional modes are unintentionally excited from the outset. This is because, following relaxation, the equilibrium state of an isentropic monatomic sphere is not modelled exactly, due to the smoothing inherent in SPH. In particular, the density is underestimated near the centre and near the boundary. (This can be improved by increasing ${\cal N}_{_{\rm NEIB}}$ and ${\cal N}_{_{\rm TOT}}/{\cal N}_{_{\rm NEIB}}$, but that requires extra computational resource.) Furthermore, the fundamental mode is excited with finite amplitude, but the eigenfunction is derived on the assumption of infinitesimal amplitude. (This can be moderated by adopting a lower value for $A$, but the test is not then relevant to real simulations of star formation, where ultimately we are concerned with non-linear perturbations.)

Setting aside the effect of other modes present in the initial conditions, the subsequent decay of the fundamental mode is in general due to two effects. First, the oscillation energy may be dissipated by artificial viscosity. The dissipation of energy due to artificial viscosity occurs wherever two neighbouring SPH particles approach one another. Second, the oscillation energy may be transferred to other modes. This occurs wherever local density or velocity fluctuations are created by the discrete nature of the SPH particles, or by the `peculiar velocities' of individual SPH particles. Both effects occur more rapidly for larger values of $\Delta{\cal N}_{_{\rm NEIB}}$. (They also occur more rapidly for smaller values of ${\cal N}_{_{\rm TOT}}$ and smaller values of ${\cal N}_{_{\rm NEIB}}$, but these parameters are not varied here.)

When $\Delta{\cal N}_{_{\rm NEIB}}=0$, the neighbour list of an SPH particle changes very seldom, and -- when it does -- very little. Therefore the acceleration experienced by the particle varies very slowly, and the velocity field is very smooth. The upshot is that neighbouring particles only approach one another very slowly, and the rate of dissipation due to artificial viscosity is low. Notwithstanding the slow rate of dissipation, there are small fluctuations in density and velocity, and these feed energy into other modes, so that the fundamental mode decays (see Figs. 1 \& 2).

As $\Delta{\cal N}_{_{\rm NEIB}}$ is increased, the neighbour list of an SPH particle changes more frequently, and more abruptly. Therefore the acceleration experienced by the particle varies in a more idiosyncratic manner, and the velocity and density fields are more noisy. The upshot is that neighbouring particles often approach one another more rapidly, and the rate of dissipation due to artificial viscosity is therefore higher. In addition, the noisy velocity and density fields are very effective at feeding energy into other modes, so that the fundamental mode decays more rapidly (see Figs. 1 \& 2).

In principle, the number of neighbours can change by $2\Delta{\cal N}_{_{\rm NEIB}}$ at each timestep, from ${\cal N}_{_{\rm NEIB}}-\Delta{\cal N}_{_{\rm NEIB}}$ to ${\cal N}_{_{\rm NEIB}}+\Delta{\cal N}_{_{\rm NEIB}}$ or vice versa. Thus with ${\cal N}_{_{\rm NEIB}}=50$ and $\Delta{\cal N}_{_{\rm NEIB}}=10$, the number of neighbours can change from 40 to 60 or vice versa. In practice such large changes are unlikely, but it is still the case that increasing $\Delta{\cal N}_{_{\rm NEIB}}$ results in increased fluctuations in the neighbour lists. In particular, particles in condensing regions tend to have ${\cal N}$ fluctuating between $\sim{\cal N}_{_{\rm NEIB}}$ and $\sim({\cal N}_{_{\rm NEIB}}+\Delta{\cal N}_{_{\rm NEIB}})$, whilst particles in expanding regions tend to have ${\cal N}$ fluctuating between $\sim{\cal N}_{_{\rm NEIB}}$ and $\sim({\cal N}_{_{\rm NEIB}}-\Delta{\cal N}_{_{\rm NEIB}})$.

In Table 1, we record, for each value of $\Delta{\cal N}_{_{\rm
    NEIB}}$ (column 1), the e-folding time of the amplitude of the fundamental mode, $T_{_A}$, in terms of its period, $P_{_0}=3.7t_{_{\rm FF}}$ (column 2); the oscillation energy left after ten periods, ${\cal E}_{_{10}}$, as a fraction of its initial value, ${\cal E}_{_0}$ (column 3); and the simulation time to evolve the oscillating gas-sphere for 10 periods on eight 2.2Ghz Opteron CPUs each with 8GB memory (column 4). The oscillation energy is given by
\begin{eqnarray}
{\cal E}&\;=\;&{\cal K}\,+\,({\cal T}-{\cal T}_{_0})\,+\,(\Omega-\Omega_{_0})\,,
\end{eqnarray}
where ${\cal T}$ is the thermal energy, and again the subscript 0 indicates the unperturbed equilibrium state.

We note that the oscillation energy decays rather slowly due to dissipation, even with $\Delta{\cal N}_{_{\rm NEIB}}=10$. This is because the oscillations have low amplitude, and therefore the relative velocities between neighbouring particles are always very subsonic. Not only are the initial amplitudes low, but in the cases with high $\Delta{\cal N}_{_{\rm NEIB}}$ the amplitudes decay rapidly due to numerical diffusion. In other words, when $\Delta{\cal N}_{_{\rm NEIB}}$ is high, the rate of dissipation is reduced because diffusion rapidly spreads the oscillation energy amongst many different modes and thereby reduces even further the relative velocities between neighbouring particles. This is reflected in the results presented in Fig. 2 and the third column of Table 1. Because the decay of the fundamental mode is largely due to numerical diffusion, the oscillation energy only falls by a few percent after ten periods (see the third column of Table 1), whereas the amplitude of the variation in kinetic energy falls much more rapidly. The variation in kinetic energy eventually disappears, not because the kinetic energy itself disappears, but because numerical diffusion transfers oscillation energy to other modes with different periods and different phases. Consequently the oscillation energy becomes thermalised, and ${\cal K}$ is finite but approximately constant.

The simulations presented here have been continued for 100 periods. In this limit, there are so many modes excited, with so many different phases, that the gas-spheres become virialised with
\begin{eqnarray}
2{\cal K}\,+\,2{\cal T}\,+\,\Omega&\;\simeq\;&0\,;
\end{eqnarray}
${\cal K}$ is still finite.

\section{Conclusions}\label{SEC:CON}

From the plots in Figs. 1 and 2, and the above discussion, we infer that the results obtained with $\Delta{\cal N}_{_{\rm NEIB}}=0$ are far more reliable than those obtained with $\Delta{\cal N}_{_{\rm NEIB}}=10$, and significantly more reliable even than those obtained with $\Delta{\cal N}_{_{\rm NEIB}}=2$, both in terms of having less dissipation and -- more importantly -- in terms of having less numerical diffusion.

Setting $\Delta {\cal N}_{_{\rm NEIB}}=0$ also requires little extra computation. For example, to follow 10 oscillation periods with $\Delta {\cal N}_{_{\rm NEIB}}=0$ takes only 17\% longer than with $\Delta {\cal N}_{_{\rm NEIB}}=10$. Moreover, 7\% of this increase is due to the fact that with $\Delta {\cal N}_{_{\rm NEIB}}=0$ the sphere continues to oscillate with a significant amplitude after 10 periods, and therefore the timestep remains short. When allowance is made for this, the real cost of reducing $\Delta {\cal N}_{_{\rm NEIB}}$ from 10 to 0 is only a 10\% increase in computation.

Therefore our principal conclusions are (i) that $\Delta{\cal N}_{_{\rm NEIB}}=0$ should be the default option for SPH codes which adapt smoothing lengths in this way; and (ii) that the Lucy test provides a very useful way of evaluating the fidelity of codes used in simulations of star formation.

We emphasise that we have not set out to reproduce as accurately or economically as possible acoustic oscillations of a self-gravitating isentropic sphere in the fundamental mode. We have simply demonstrated how the results produced using a standard SPH code, with a modest number of particles (${\cal N}_{_{\rm TOT}}=5,895$) depend on $\Delta{\cal N}_{_{\rm NEIB}}$. There are adjustments to SPH which will improve (extend) the timescales for dissipation and numerical diffusion in the present simulation. For example, using standard artificial viscosity with $\alpha=0$ and $\beta=0.1$ increases the e-folding time for the fundamental mode to $\sim 60$ oscillation periods, but at the same time compromises the shock-capturing ability of the code so that it can not then be used for simulations in which shocks are likely. Similarly, the e-folding time for the fundamental mode can be extended by increasing ${\cal N}_{_{\rm NEIB}}$ or ${\cal{\cal N}_{_{\rm TOT}}/\cal N}_{_{\rm NEIB}}$, but this must inevitably be at the expense of resources which are needed elsewhere, {\it viz.} to maximise the extent and/or duration of a simulation.

The useful duration of the present simulation can be identified with the e-folding time of the fundamental mode, which with $\Delta{\cal N}_{_{\rm NEIB}}=0$ is 13.6 oscillation periods, but with $\Delta{\cal N}_{_{\rm NEIB}}=10$ is only 3.1 oscillation periods\footnote{In fact, even 3.1 oscillation periods is an overestimate of the e-folding time in the simulation with $\Delta{\cal N}_{_{\rm NEIB}}=10$, since this case is so noisy that numerical diffusion actually re-invigorates the fundamental mode, but with a phase-shift relative to the initial oscillation!}.

Adaptive Mesh Refinement codes used for simulating star formation should also aim to reproduce or improve upon the results produced here, using comparable computational resources. {\em In addition}, they should do so for a gas-sphere which moves at constant velocity relative to the underlying Cartesian grid, in order to match the Lagrangian advantages of SPH.

\begin{acknowledgements}
REA acknowledges the support of a PPARC Studentship (PPA/S/S/2004/03981). SPG acknowledges the support of a UKAFF Fellowship. APW acknowledges the support of a PPARC Rolling Grant (PPA/G/O/2002/00497). We are very grateful to Alfred Gautschy for supplying us with the eigenfunction for the fundamental radial mode of an isentropic, monatomic, self-gravitating gas sphere, and to the referee Leon Lucy for useful comments on the first version.
\end{acknowledgements}


\end{document}